\input iopppt
\pptstyle

%
\def\({(eq:}
\def\HS{1}
\def\Sch{2}
\def\Bondi{3}
\def\Emch{4}
\def\Penrose{5}
\def\TY{6}
\def\And{7}
\def\Gil{8}
\def\Gae{9}
\def\Mich{10}
%
%
\def\plane{1}
\def\ricciflat{2}
\def\Killing{3}
\def\Killingeq{4}
\def\symmetric{5}
\def\signature{6}
\def\homo{8}
\def\equi{9}
\def\equation{10}
\def\inveq{11}
\def\Taylor{12}
\def\scale{13}
\def\mini{28}
\def\tensor{23}
\def\nteq{21}
\def\redEinstein{25}

\def\AFT{11}
\title{Gravitational waves: just plane 
symmetry}
\author{C G 
Torre}
\address{Institute for 
Theoretical Physics, University of California, Santa 
Barbara, CA 93106, USA}
\address{Department of Physics, Utah State 
University, Logan, UT 84322-4415 USA\footnote*{Permanent 
address.}} 

\abs
We present some remarkable properties of the symmetry group for 
gravitational plane waves. Our main observation is that metrics 
with plane wave symmetry satisfy every 
system of generally covariant vacuum field equations {\it except} 
the 
Einstein equations. 
 The proof uses the homothety admitted by metrics with plane wave 
symmetry and the scaling behavior of generally covariant field 
equations.
We 
also discuss  a mini-superspace description of spacetimes with 
plane wave symmetry.
\endabs
\bigskip\noindent
July 15, 1999



\section{Introduction}

We would like to present some remarkable properties of the symmetry 
group of gravitational plane waves. Our main observation is that 
metrics with plane wave symmetry satisfy every system 
of generally covariant vacuum field equations {\it except} 
the Einstein 
equations. Put slightly differently, the only non-trivial field 
equations that can be imposed on a metric with plane wave 
symmetry 
are equivalent to the vacuum Einstein equations. Let us call a 
Ricci-flat metric that admits plane wave 
symmetry a {\it 
gravitational plane wave}. Gravitational plane waves satisfy 
every generally covariant system of vacuum field equations and 
so can be 
called 
``universal solutions'' of the Einstein equations.

Related results can be found in 
work by Horowitz and Steif [\HS]. In 
[\HS] it is shown that metrics which  (1) are Ricci-flat, (2) 
admit 
a constant (null) vector field\footnote*{
Metrics satisfying (1) and (2) are the ``plane fronted'' waves, 
a class of vacuum metrics 
that includes the gravitational 
plane waves.}, satisfy all other field equations that are
symmetric rank two tensors covariantly constructed
from scalar invariants and 
polynomials in the curvature and their covariant derivatives.  
By contrast, our hypothesis 
is only that the metric admits  plane wave symmetry, and the 
field equations being considered are any symmetric rank two 
tensors which are covariantly constructed as 
smooth, local functions of the metric and its 
derivatives to any order (with no polynomiality or analyticity 
assumptions). 
Our proof of the 
``universality'' of metrics with plane wave symmetry is based 
upon 
an interplay between the  homothety admitted by metrics with 
plane wave symmetry and the scaling properties of generally 
covariant field equations. The proof can be viewed, to some 
extent, as a generalization of
an idea  of Schmidt [\Sch], who used the homothety 
admitted 
by gravitational plane waves
to show that all generally covariant 
scalars are constant. Our 
arguments are valid in 
any spacetime dimension and can also be 
used to show that a number of other covariantly constructed 
tensor 
fields must vanish when evaluated on metrics with plane wave 
symmetry. 
  
In the next section we specify what we mean by ``plane wave 
symmetry'', and construct the most general metric with that 
symmetry. We work in an arbitrary number of spacetime dimensions. 
It follows 
easily that metrics with plane wave symmetry always admit a 
continuous homothety. In \S 3 we define ``generally covariant 
field equations'' and indicate their behavior under scaling 
of the metric.
 In \S 4 we combine the results of \S 2 and \S3 to show 
that metrics with plane wave symmetry satisfy every system of 
field equations except the Einstein equations. In \S 5 we mention 
some other results along these lines.  We also comment on the 
construction of a mini-superspace description of spacetimes with 
plane wave symmetry.

\section{ Plane wave symmetry}
The spacetime corresponding to a gravitational plane wave 
[\Bondi,\Emch] can be defined as
  the 
manifold $M={\bf R}^n$, 
with standard global
coordinates $x^\alpha=(u,v,x^i)$, $i=1,2,\dots, n-2$, and metric
$$
g_{\scriptscriptstyle plane} = -2 du\otimes dv + \delta_{ij} 
dx^i\otimes dx^j
+ f_{ij}(u)x^i x^j du\otimes du,
\en
$$
where $f_{ij}=f_{ji}$ are any smooth functions of the null 
coordinate $u$ such that
$$
f_i^i \equiv \delta^{ij} f_{ij} = 0.
\en
$$
Condition 
(\ricciflat) renders $g_{\scriptscriptstyle plane}$ Ricci-flat.
Henceforth we raise and lower the Latin indices using the 
Kronecker delta as 
in (\ricciflat). The wave profile (amplitude, 
polarization, {\it etc.}) is determined by the choice of 
the 
functions $f_{ij}$ since the components of the Weyl 
tensor in the $(u,v,x^i)$ chart are proportional to the functions 
$f_{ij}$. 
For any smooth choice of wave profile $f_{ij}$, the plane wave 
spacetime is geodesically complete, and stably causal [\Emch]. Thus 
the 
plane wave spacetimes provide a relatively rare class of examples 
of 
non-singular, causally tame vacuum solutions with a rather 
simple physical interpretation. One somewhat pathological feature 
of these spacetimes is that 
they are not globally hyperbolic, as was first noticed by Penrose
[\Penrose].

The metric (\plane) admits a $(2n-3)$-dimensional 
group $G$ of isometries 
generated by 
 the vector fields
$$
{\partial\over \partial v}\quad{\rm and}\quad
Y = S^i(u) {\partial\over\partial x^i} + S_i^{\prime}(u) 
x^i{\partial\over\partial v},
\en
$$
where a prime denotes a $u$ derivative, and  $S^i(u)$ is any 
smooth solution of the linear system
$$
S^{i\prime\prime} = f^i_j S^j.
\en
$$
The solution space to (\Killingeq) is $2(n-2)$-dimensional, being 
labeled by the 
initial data $S^i(u_0)$ and $S^{i\prime}(u_0)$. Thus one can view 
$Y$, as defined in (\Killing), as representing $2(n-2)$ Killing 
vector fields, corresponding to any choice of basis for the 
solution
space to (\Killingeq).

Basic 
properties of 
the isometry group of gravitational plane waves 
can be found in [\Bondi,\Emch]. We mention here some 
properties of the isometry group that will feature in what 
follows.

The group orbits are the null hypersurfaces $u={\rm constant}$.  
Any function invariant under the plane wave symmetry group is a 
function of $u$ only. We call such functions {\it $G$-invariant 
functions}. 
The components of the Killing vector fields $Y$ depend 
upon the invariant $u$. This dependence cannot be 
removed by a coordinate transformation and reflects the fact 
that, roughly speaking, the group action varies from orbit to 
orbit.  More precisely, the orbits of the symmetry group, 
while diffeomorphic, are 
distinct as homogeneous spaces. 
Related to this is the fact that 
the transformation group generated by 
the 
Killing vector fields depends upon the choice of the functions 
$f_{ij}$ in (\plane) so 
that, strictly speaking, different plane wave spacetimes have 
different symmetry transformations (although the abstract 
$(2n-3)$-dimensional Lie group is the same for all choices 
of the wave profile). The 
dependence of the symmetry transformations on the choice of 
 wave 
profile is through the functions $S^i(u)$, which are functionals 
of $f_{ij}(u)$ via (\Killingeq).  

Our first task is to find the general form of a metric admitting a 
plane wave symmetry group. We do this by considering a fixed (but 
arbitrary) set of functions $f_{ij}(u)$ (satisfying 
(\ricciflat), although 
this is not essential) and then defining the 
vector fields $Y$ as in (\Killing), (\Killingeq). We then find 
all 
metrics with Lorentz signature whose Lie derivative along the 
vector fields $Y$ and ${\partial\over\partial v}$ vanish. 
The general form of a metric $g$ 
admitting plane wave symmetry is then found to be
$$
g = \alpha g_{\scriptscriptstyle plane} + \beta du\otimes 
du,
\en
$$
where $\alpha$ and $\beta$ are $G$-invariant functions, 
{\it i.e.}, 
$\alpha=\alpha(u)$, $\beta=\beta(u)$, and
$$
 \alpha > 0
\en
$$
is required to give  $g$ the Lorentz signature. If 
we drop  condition (\signature), then (\symmetric) is the 
general form of a symmetric rank-2 tensor field invariant under 
the plane wave symmetry group characterized by $f_{ij}$.

Next, we point out  that the $G$-invariant metric 
 given in (\symmetric), (\signature) admits a 
 continuous homothety
for any choice of the $G$-invariant functions $\alpha$ and 
$\beta$. 
This means that there exists a one-parameter family of 
diffeomorphisms  
$\Psi_s\colon M\to M$ such 
that
$$
\Psi_s^* g = s^2 g,\quad s>0.
\en
$$
The homothety is given by the transformation
$$
\eqalign{
&u\to u\cr
&v \to s^2 v + {1\over 2}(1-s^2) 
\int {\beta(u)\over \alpha(u)} \, du\cr
&x^i\to s x^i.}
\en
$$
Note that the homothetic transformation preserves the orbits of 
the plane wave symmetry group, that is, $u$ is invariant under 
the 
homothety.

We summarize this section as follows.

\proclaim {Definition 1}
A tensor field on ${\bf R}^4$ admits a {\sl plane wave symmetry} 
with wave profile $f_{ij}$
if it is invariant under the group of diffeomorphisms generated 
by 
${\partial \over \partial v}$ and $Y$, given in (\Killing), 
(\Killingeq).
\endproclaim

\proclaim {Proposition 1}
If a symmetric tensor field $g$ of type $\left({}^0_2\right)$ has 
plane wave symmetry, then it takes the form (\symmetric) for some 
choice of the $G$-invariant functions $\alpha$ and $\beta$. If 
$g$ is a 
Lorentz metric with plane wave symmetry then it takes the form 
(\symmetric) with $\alpha>0$.
\endproclaim

\proclaim {Proposition 2}
If a metric has plane wave symmetry, then it admits a continuous 
homothety which preserves the orbits of the plane wave 
symmetry group.
\endproclaim

\section{Generally Covariant Field Equations}

We now characterize the set of field equations that we want to 
consider. We will consider field equations for a metric 
that take the form of an equality between 
``generally covariant'' tensor fields on a given manifold $M$. 
Such tensor fields   are  often 
called just ``tensor fields'', or ``natural tensor fields'', or 
``invariant tensor fields'', or ``metric concomitants''. 
Whatever the name, the point is that 
such tensor fields 
are globally defined by the metric, with no other structures 
being needed. 
If the manifold $M$ is orientable (as it is for the plane wave 
spacetimes), it is sensible to fix an orientation and to enlarge 
the class of 
generally covariant field equations by allowing the orientation  
of 
the manifold to be used  in their construction (via the volume 
form defined by the metric).  All 
the results that follow are valid with or without the use of an 
orientation on $M$.  The precise implementation of our general 
covariance criteria 
is as follows.

\proclaim {Definition 2}
A {\sl generally covariant
tensor} of type 
$\left({}^p_q\right)$ built from a metric, denoted $T$,  
is a mapping that assigns to each metric $g$ a tensor 
field $T[g]$
of type 
$\left({}^p_q\right)$ on any 
manifold $M$. This rule must be 
smooth 
and local, that is, in any chart about any point $x\in M$, the 
components of $T[g]$ are smooth functions 
of 
the components of the metric and their derivatives (to some 
finite 
order) at $x$.  Finally, we require for 
any 
(orientation-preserving) diffeomorphism, $\phi\colon M\to M$, 
that
$$
T[\phi^* g] = \phi^* T[g].
\en
$$
\endproclaim

Because we are considering metric field theories only, we 
have restricted our notion of generally covariant tensors to 
those that are constructed from a metric. If other 
 fields ({\it e.g.,} electromagnetic) were to be 
 considered, we would of course enlarge the definition of 
 generally covariant tensors accordingly. It is a 
standard result [\TY] that generally covariant tensors can always 
be 
constructed  as smooth functions  of the metric, 
the volume form of the metric (in the orientation-preserving 
case),  the curvature tensor, 
and covariant derivatives of the curvature tensor to some finite 
order, all of which are examples of generally covariant tensors.  
Note that we use the symbol $T$ to denote the mapping from 
metrics to tensor fields, and we use the symbol $T[g]$ to 
denote a specific tensor field on $M$ defined by applying 
the rule $T$ to a given metric tensor field $g$ on $M$.

\proclaim {Definition 3}
A set of {\sl generally covariant field equations} for a metric 
is defined by partial differential equations of the form
$$
T[g] = 0,
\en
$$
where
$T$ is a generally covariant symmetric tensor of type 
$\left({}^0_2\right)$.
\endproclaim

The first result we need is  that generally 
covariant tensor fields inherit the symmetries of the metric used 
to construct them. This follows directly from the equivariance 
requirement  (\equi) 
when  $\phi$ is an isometry of the 
metric ({\it i.e.}, $\phi^*g = g$).
In particular,  let $T$ be a
generally covariant symmetric tensor of type 
$\left({}^0_2\right)$, and let $g$ be a metric on $M$ 
with plane wave 
symmetry, then the tensor field $T[g]$ on $M$ 
takes the form (\symmetric). With a simple redefinition of the 
functions $\alpha$ and $\beta$, we have the following result.

\proclaim {Proposition 3}
Let $T$ be a generally covariant symmetric tensor of type 
$\left({}^0_2\right)$, and $g$ a metric with plane wave symmetry, 
then there exist $G$-invariant functions $\rho$ and $\sigma$ 
such that
$$
T[g] = \rho g + \sigma du\otimes du.
\en
$$
\endproclaim

We remark that the functions $\rho$ and $\sigma$, while 
showing up as functions on $M$ in (\inveq), 
also can be viewed 
 as generally covariant scalar fields, that is, 
 they are obtained 
by evaluating generally covariant tensors of 
type $\left({}^0_0\right)$ 
on $g$.

The other result we need concerns the behavior of generally 
covariant tensors with respect to scaling of the metric. 
From the 
work of Anderson [\And] and Gilkey [\Gil] 
we have the following result.

\proclaim {Proposition 4}
Let $T$ be a generally covariant tensor of type 
$\left({}^p_q\right)$ and let $g$ be any metric tensor field, 
then $T[g]$ can be written as
$$
T[g] = T_0[g] + T_1[g] + T_2[g] + \dots + T_N[g] + R_N[g],
\en
$$
where each of $T_i$, $i=1,2,\dots, N$, and $R_N$ are generally 
covariant tensors of type $\left({}^p_q\right)$ 
that enjoy the scaling behavior:
$$
\eqalign{
T_j[s^2 g] &= s^{q-p-j} T_j[g],\cr
R_N[s^2g] &= {\cal O}(s^{q-p-N-1}).}
\en
$$
\endproclaim
Here the notation $A={\cal O}(s^{r})$ means that $s^{-r}A$ 
has a limit as $s\to 0$.
Using the results of [\And] it is not hard to show that, 
when $T$ is 
symmetric and of type $\left({}^0_2\right)$, 
$$
\eqalignno{
&(T_0)_{\mu\nu} = a g_{\mu\nu},\en\cr
&(T_1)_{\mu\nu} = 0,\en\cr
&(T_2)_{\mu\nu}=
b R_{\mu\nu} + c R g_{\mu\nu},\en}
$$
where $a$, $b$, $c$ are constants, $R_{\mu\nu}$ is the Ricci 
tensor 
and $R$ is the scalar curvature.

\section{Universality}

We now show that metrics with plane wave symmetry are universal 
in 
the sense that they satisfy ``almost all'' generally covariant 
field equations.  Given a set of generally covariant field 
equations (\equation),  we can expand 
$T[g]$ as in (\Taylor); each term in the expansion is a natural 
tensor field with scaling 
behavior (\scale). We suppose that $g$ is a metric with 
plane wave symmetry. Using the homothety $\Psi_s$, given in 
(\homo), we have 
that
$$
\Psi_s^*T_j[g] = T_j[\Psi_s^*g] = T_j[s^2 g] =  s^{2-j} T_j[g],
\en
$$
where the first equality comes from (\equi) and the last equality 
comes from (\scale). On the other hand, Proposition 3 allows us 
to conclude that there exist $G$-invariant functions $\rho$ and 
$\sigma$ such 
that 
$$
\Psi_s^*T_j[g] = \Psi^*_s\left(\rho g + \sigma du\otimes 
du\right) = s^2\rho  g + \sigma du\otimes du,
\en
$$
where we used the fact that the 1-form $du$ is invariant under 
the homothety as is any $G$-invariant function.
Therefore, for all $s>0$,
$$
s^2\rho  g + \sigma du\otimes du=s^{2-j}
\left( \rho  g + \sigma du\otimes du\right),
\en
$$
which implies that either $j=2$ and $\rho=0$, or that $j=0$ and 
$\sigma=0$, or that $\rho=\sigma=0$. 
Similarly, it follows that $R_N[g]=0$ for $N>2$. Thus, 
at 
most, 
$$
T[g] = T_0[g] +  T_2[g].
\en
$$
Furthermore, either from direct computation or by an application 
of a scaling argument analogous to that just described (see 
Theorem 2, below), it is 
easily seen that the scalar curvature vanishes 
for any metric with plane wave symmetry. Therefore we have the 
following result.

\proclaim {Theorem 1}
Let $g$ be a metric with plane wave symmetry, 
and suppose that $T$ is a generally covariant symmetric 
tensor of type $\left({}^{0}_{2}\right)$. 
Then $T[g]$ is a linear combination of $g$ and the Ricci 
tensor of 
$g$. 
\endproclaim

Thus the only generally covariant field equations that 
are {\it 
not} automatically satisfied by a metric with plane wave 
symmetry 
are of the form
$$
a g_{\mu\nu} + b  R_{\mu\nu} = 0,
\en
$$
with $a$ and $b$ constants. More explicitly, if 
$$
g=\alpha(u) g_{\scriptscriptstyle plane} + 
\beta(u) du\otimes du,
\en
$$
then the field equations (\nteq) are
$$
 a [\alpha g_{\scriptscriptstyle plane} + \beta du\otimes 
du] + b[{\alpha^{\prime\prime}\over\alpha} - {3\over 
2}\left({\alpha^{\prime}\over\alpha}\right)^{2}]du\otimes 
du=0.
\en
$$
From (\tensor) it is easy to see that the Ricci 
tensor of a metric 
with plane wave symmetry is 
proportional to $du\otimes du$ (this is the case $j=2$, 
$\rho=0$ 
mentioned earlier), so (\nteq) has no solutions 
unless $a=0$, {\it i.e.}, the cosmological constant 
must vanish. 

\proclaim {Corollary}
Any  constraints that can be placed 
by generally covariant field equations
upon a metric with plane wave symmetry 
are equivalent to the vacuum Einstein equations with 
vanishing cosmological constant.
\endproclaim

\section{Remarks}

\noindent{\bf (1)}
The ability of the plane wave symmetric spacetimes to 
satisfy so many field equations is  reminiscent of the 
idea of ``critical solutions'' [\Gae]. These are field 
configurations which  (i) are invariant under some symmetry 
group $G$, and (ii) are critical points for any $G$-invariant 
action functional. The existence of critical solutions can, 
in many instances, be viewed as an infinite-dimensional 
analog of Michel's theorem [\Mich], which states that a 
point is a critical point for all $G$-invariant functions 
if and only if it is ``isolated in its 
stratum''. It 
is not clear if one can view gravitational plane waves as 
critical solutions of this type, if only because the 
metrics with plane wave symmetry are {\it not} critical 
points of the Einstein-Hilbert action. On the other 
hand, by working in four dimensions and  
by restricting the form of the wave profile, it is 
possible to enlarge the plane wave symmetry group such that the 
resulting group invariant metrics {\it are} automatically 
Ricci-flat. For example, let $\kappa$ be a constant,  let
$$
f_{ij} = \kappa\left(\matrix{ \cos(2\kappa u) &\sin(2\kappa 
u)\cr
\sin(2\kappa u) &-\cos(2\kappa u)}\right),
$$
and adjoin to the 5 generators in (\Killing) the vector 
field
$$
Z = {\partial\over\partial u} -\kappa 
(y{\partial\over\partial x} - x{\partial\over\partial y}).
$$
The commutator of $Z$ with 
${\partial\over \partial v}$ vanishes, and it is a 
straightforward exercise to check 
that the commutator of $Z$ with the 4 independent 
vector fields 
defined by $Y$ is a 
linear combination of those vector fields. 
The enlarged group is thus six-dimensional with four 
dimensional orbits; the resulting group invariant 
metrics define homogeneous
spacetimes with plane wave symmetry
(see the article by Ehlers and 
Kundt [\Bondi] for another example). Using the enlarged
symmetry group, the general form 
of the group-invariant metric is now
$$
\eqalign{
g&=
a \big\{-2 du\otimes dv + dx \otimes dx + dy \otimes  dy 
\cr
&+ [b + \kappa \cos(2\kappa 
u)(x^{2} - y^{2}) + 2\kappa\sin(2\kappa u) xy)] 
du\otimes  du\big\},}
\en
$$
with $a$ and $b$ constants. This metric is Ricci-flat for 
all values of $a$, $b$ and $\kappa$. 
Therefore, these 
spacetimes will solve all generally covariant vacuum 
field equations 
by virtue of their symmetry, and perhaps can be understood 
via Michel's theorem. In any case, it is worth 
noting in this regard that  gravitational 
plane waves satisfy {\it all} 
generally covariant 
vacuum field equations; the vast majority of these 
equations are 
not derivable from a local variational principle.

\bigskip\noindent
{\bf (2)} The same sort of arguments as used in \S 4 can be 
used to 
investigate the behavior of 
 generally covariant tensor fields of other types. 
For example, it is not hard to establish the following.

\proclaim {Theorem 2}
Let $g$ be a metric with plane wave symmetry. 
{\bf (i)} All generally 
covariant scalar fields are constant when evaluated 
on $g$. Here 
``constant'' means as 
a 
function on $M$ and as a functional of metrics with 
plane wave 
symmetry. {\bf (ii)} All generally 
covariant 1-forms, 2-forms, and 3-forms vanish when 
evaluated on 
$g$. {\bf (iii)} All generally covariant 4-forms are 
constant multiples of  
the volume form of $g$.
\endproclaim

\bigskip\noindent
{\bf (3)} It is possible to consider a
``mini-superspace'' description of spacetimes 
admitting the plane 
wave symmetry group. The symmetry group is 
defined by a choice of 
wave profile as discussed in \S2. The 
mini-superspace $\cal S$ is 
defined in terms of the space of metrics 
of the form
(\symmetric), so that points in $\cal S$ 
are specified by 
the values of
$\alpha$ 
and $\beta$;  the mini-superspace is 
two-dimensional. As we have 
seen, the only field equations that can be 
imposed are the vacuum 
Einstein equations, which take the form
$$
\alpha^{\prime\prime} 
-{3\over2}{1\over \alpha}\alpha^{\prime 2} 
=0.
\en
$$
Evidently, the variable $\beta$ is 
``pure gauge'' and completely 
drops out of 
the field equations. This reflects the fact 
that the value of 
$\beta$ can be varied at will by making a coordinate 
transformation of the form
$$
v\to v + \Lambda(u).
\en
$$
We might as well drop $\beta$ from the mini-superspace. 
Defining $q$ via
$$
\alpha = {1\over q^2},
\en
$$

the field equations take the elementary form
$$
q^{\prime\prime} = 0.
\en
$$
Note that the reduced equations of motion,
(\redEinstein) or (\mini), are not invariant with 
respect to reparametrizations of the ``time'' 
$u$, contrary to 
what might be 
expected. This is due to the fact that the 
symmetry generators 
(\Killing) depend explicitly upon the $u$ 
coordinate, so that the 
symmetry group defining the reduced equations 
of motion is sensitive to 
reparametrizations of $u$.

Because the  scalar curvature vanishes when 
evaluated on metrics 
with plane wave symmetry, one  cannot simply insert the 
general metric with plane wave symmetry into the
Einstein-Hilbert Lagrangian to obtain a reduced Lagrangian 
describing the dynamics on the plane wave
mini-superspace.\footnote{\ddag}{For necessary and sufficient 
conditions 
 on a symmetry group $G$ such that one {\it can} make the 
symmetry 
reduction at the level of the Lagrangian, see 
[\AFT] and references 
therein.} 
Nevertheless, the equation of motion (\mini) 
obviously admits a Lagrangian, 
so one can view a gravitational plane wave (with a given wave 
profile) as an autonomous
Hamiltonian system with one degree of freedom.

\ack
Many thanks to Ian Anderson and Mark Fels for comments and 
complaints. I would also like to acknowledge helpful Usenet 
correspondence with Chris Hillman and Robert Low.
This work was supported in part by National Science 
Foundation grants PHY97-32636 and 
PHY94-0719. 

\numreferences

\numrefjl{[\HS]}{Horowitz G and Steif A 1990}{\PRL}{64}{260}

\numrefbk{[\Sch]}{Schmidt H 1996}{New frontiers in 
gravitation}{ed. Sardanashvily G (Palm Harbor, Fla. : Hadronic 
Press) p~337}

\numrefjl{[\Bondi]}{Bondi H, Pirani F, Robinson I 1959}{Proc. 
Roy. Soc. London}{A251}{519}

\numrefbk{}{Ehlers J and Kundt W 1962}{Gravitation}{ed L. 
Witten (Chichester: Wiley) p~49}

\numrefjl{[\Emch]}{Ehrlich P and Emch G 1992}{Rev. Math. 
Phys.}{4}{163}

\numrefjl{[\Penrose]}{Penrose R 1965}{\RMP}{37}{215}

\numrefbk{[\TY]}{Thomas T 1934}{Differential Invariants of 
Generalized Spaces}{(London: Cambridge University Press)}

\numrefjl{[\And]}{Anderson I 1984}{Ann. Math.}{120}{329}

\numrefjl{[\Gil]}{Gilkey P 1978}{Adv. Math.}{28}{1}

\numrefjl{[\Gae]}{Gaeta G and Morando P 1997}{Ann. 
Phys.}{260}{149}

\numrefjl{[\Mich]}{Michel L 1971}{C. R. Acad. Sci. 
Paris}{A272}{433}

\numrefjl{[\AFT]}{Palais R 1979}{Commun. Math. Phys.}{69}{19}

\numrefjl{}{Anderson I and Fels M 1997}{Am J. Math.}{119}{609}

\numrefjl{}{Torre C 1999}{Int. J. Th. Phys.}{38}{1081}

\bye